\documentclass[aps,12pt,preprintnumbers,nofootinbib,superscriptaddress]{revtex4}
\usepackage{graphicx}
\usepackage{amssymb,epsfig,amsmath}
\begin{document}

\title{\Large{\bf Phase Transition of RN-AdS Black Hole with Fixed Electric Charge and Topological Charge  } }

\author{Shan-Quan Lan}
\email[ ]{shanquanlan@126.com}
\affiliation{%
Department of Physics, Lingnan Normal University, Zhanjiang, 524048, Guangdong, China}

\date{\today}

\begin{abstract}

Phase transition of RN-AdS black hole is investigated from a new perspective. Not only the cosmological constant is treated as pressure, but also the spatial curvature of black hole is treated as topological charge $\epsilon$. We obtain the extended thermodynamic first law from which the mass is naturally viewed as enthalpy rather than internal energy. In canonical ensemble with fixed topological charge and electric charge $Q$, interesting van der Waals like oscillatory behavior in $T-S$, $P-V$ graphs  and swallow tail behavior in $G-T$, $G-P$ graphs are observed. By applying the Maxwell equal area law and analysing the gibbs free energy, we obtain analytical phase transition coexistence curves which are consistent with each other. The phase diagram is four dimensional with $T,P,Q,\epsilon$.

\end{abstract}
\maketitle
\newpage

\date{\today}

\section{Introduction}

Black hole is a simple object which can be described by only a few physical quantities, such as mass, charge, angular momentum, etc. While, it is also a complicate thermodynamic system. Since the discovery of black hole's entropy\cite{Bekenstein:1973ur}, the four thermodynamic law\cite{Bardeen1973}, and the Hawking radiation\cite{hawking1975} in 1970s, thermodynamic of black hole has become an interesting and challenging topic. Especially, in the anti-de Sitter (AdS) space, there exists Hawking-Page phase transition between stable large black hole and thermal gas\cite{Hawking}. Due to the AdS/CFT correspondence\cite{Maldacena,Gubser,Witten}, the Hawking-Page phase transition is explained as the confinement/deconfinement phase transition of a gauge field\cite{Witten2}.

When the AdS black hole is electrically charged, its thermodynamic properties become more rich. In the canonical ensemble with fixed electric charge, there is a first-order phase transition between small and large black holes\cite{Chamblin,Roychowdhury,Banerjee,Chamblin2}. Increasing the temperature, the phase transition coexistence curve ends at the critical point, where the first-order phase transition becomes a second-order one. In the grand canonical ensemble with fixed temperature, there is also a critical temperature. Below the critical temperature, $\Phi(Q)$ is a single-valued function, where $Q$ is electric charge and $\Phi$ is the conjugate potential. Above the critical temperature, $\Phi(Q)$ is a multivalued function with phase transitions\cite{Chamblin2}. The phase transition behavior of AdS black hole is reminiscent to the liquid-gas phase transition in a van der Waals system.

Viewing the cosmological constant as a dynamical pressure and the black hole volume as its conjugate quantity\cite{Kastor:2009wy}, the analogy of charged AdS black hole as a van der Waals system has been further enhanced in Ref.\cite{2012JHEP07033K}. Both the systems share the same oscillatory behavior in pressure-volume  graph and swallow tail behavior in gibbs free energy- temperature (pressure) graph. What's more, they have very similar phase diagrams and have exactly the same critical exponents. The phase transition property is also investigated in temperature-entropy graph\cite{2013arXiv13053379S}. Later, this analogy has been generalized to different AdS black holes, such as rotating black holes, higher dimensional black holes, Gauss-Bonnet black holes, f(R) black holes, black holes with scalar hair, etc\cite{Gunasekaran,Hendi,Chen,ZhaoZhao,Altamirano,Cai,AltamiranoKubiznak,XuXu,Mo,zou,MoLiu,Altamirano3,
Wei,Wei2,Zhang,Moliu,Zou2,Zhao2,ZhaoZhang,XuZhang,Frassino,Zhangcai,Mirza,Kostouki,Rajagopal,Liuwang,
Hendi:2017fxp,Hendi:2016vux,2013arXiv13053379S,Lan:2015bia,Ma:2017pap,Wei:2017icx,Bhattacharya:2017hfj,
Hendi:2016usw,Kuang:2016caz,Fernando:2016sps,Majhi:2016txt,Zeng:2016aly,Sadeghi:2016dvc,Zeng:2015wtt,
Nguyen:2015wfa,Xu:2015rfa,Nie:2015zia}, where more interesting phenomena are found.

Recently, the spatial curvature of electrically charged AdS black hole is viewed as variable and treated as topological charge\cite{Tian:2014goa,Tian2018hlw} in Einstein-Maxwell's gravity and Lovelock-Maxwell theory. The authors found that the topological charge naturally arisen in holography.  What is more, together with all other known charges ( electric charge, mass, entropy), they satisfy an extended first law and the Gibbs-Duhem-like relation as a completeness. In our last paper\cite{lan1804}, when the cosmological constant is not viewed as variable, we find a van der Waals type but new phase transition relating to the topological charge. While, in this paper, we will treat both the cosmological constant and the spatial curvature as variables, then following one of their methods to derive the extended first law, from which one can see the cosmological constant is naturally viewed as pressure and the mass is viewed as enthalpy. Based on the extended first law, the black hole's phase transition property will be investigated in canonical ensemble with fixed electric charge and topological charge.

This paper is organized as follows. In Sec.\ref{sec2}, following the method in Ref.\cite{Tian2018hlw}, we will derive the extended first law in d dimensional space-time. In Sec.\ref{sec3}, by analysing the specific heat, the phase transition of AdS black hole in 4 dimensional space-time is studied and the critical point is determined. In Sec.\ref{sec4}, the van der Waals like oscillatory behavior is observed in both $T-S$ and $P-V$ graphs. Then we use the Maxwell equal area law  to obtain the phase transition coexistence curve. In Sec.\ref{sec5}, the van der Waals like swallow tail behavior is observed in $G-T$ and $G-P$ graphs, then we will obtain the phase transition coexistence curve by analysing the gibbs free energy.  Finally, we summarize and discuss our results in Sec.\ref{sec6}.

\section{the extended thermodynamic first law }
\label{sec2}

The d dimensional space-time AdS black hole solutions with maximal symmetry in the Einstein-Maxwell theory are
\begin{equation}\label{metric}
  ds^{2}=\frac{dr^{2}}{f(r)}-f(r)dt^{2}+r^{2}d\Omega^{(k)2}_{d-2},
\end{equation}
where
\begin{eqnarray}
  f(r)&=&k+\frac{r^{2}}{l^{2}}-\frac{m}{r^{d-3}}+\frac{q^{2}}{r^{2d-6}},\nonumber\\
  d\Omega^{(k)2}_{d-2}&=&\hat{g}^{(k)}_{ij}(x)dx^{i}dx^{j},\nonumber\\
  A&=&-\frac{\sqrt{d-2}q}{\sqrt{2(d-3)}r^{d-3}}dt.
\end{eqnarray}
$m,q,l$ are related to the ADM mass $M$, electric charge $Q$, and cosmological constant $\Lambda$ by
\begin{eqnarray}
  M&=&\frac{(d-2)\Omega^{(k)}_{d-2}}{16\pi}m,\nonumber\\
  Q&=&\sqrt{2(d-2)(d-3)}(\frac{\Omega^{(k)}_{d-2}}{8\pi})q,\nonumber\\
  \Lambda &=&-\frac{(d-1)(d-2)}{2l^{2}},
\end{eqnarray}
 and $\Omega^{(k)}_{d-2}$ is the volume of the ``unit" sphere, plane or hyperbola, $k$ stands for the spatial curvature of the black hole. Under suitable compactifications for $k\leq 0$, we assume that the volume of the unit space is a constant $\Omega_{d-2}=\Omega^{(k=1)}_{d-2}$ hereafter\cite{Tian:2014goa,Tian2018hlw}.

 Following Ref.\cite{Tian2018hlw}, the first law of thermodynamics can be derived. As the first law of thermodynamics is about the differential relation of every physical quantities, one can first find an equation containing these physical quantities and then differentiate it to obtain the first law of thermodynamics. Considering an equipotential surface $f(r)=c$ with fixed $c$ (here set $c=0$), we variate both sides of the equation and obtain
\begin{eqnarray}
  df(r_{+},k,m,q)=\frac{\partial f}{\partial r_{+}}dr_{+}+\frac{\partial f}{\partial k}dk+\frac{\partial f}{\partial \frac{1}{l^{2}}}d\frac{1}{l^{2}}+\frac{\partial f}{\partial m}dm+\frac{\partial f}{\partial q}dq=0,
\end{eqnarray}
where $r_{+}$ is the radius of event horizon. Noting
\begin{eqnarray}
  &\,&\partial_{r_{+}}f=4\pi T,\,\,\,\,\,\partial_{k}f=1,\,\,\,\,\,\partial_{\frac{1}{l^{2}}}f=r_{+}^{2},\nonumber\\
  &\,&\partial_{m}f=-\frac{1}{r_{+}^{d-3}},\,\,\,\,\,\partial_{q}f=\frac{2q}{r_{+}^{2d-6}},
\end{eqnarray}
we obtain
\begin{equation}
  dm=\frac{4\pi T}{d-2}dr_{+}^{d-2}+r_{+}^{d-3}dk+r_{+}^{d-1}d\frac{1}{l^{2}}+\frac{2q}{r_{+}^{d-3}}dq.
\end{equation}
Multiplying both sides with an constant factor $\frac{(d-2)\Omega_{d-2}}{16\pi}$, the above equation becomes
\begin{equation}
  dM=TdS+\frac{(d-2)\Omega_{d-2}}{16\pi}r_{+}^{d-3}dk+\frac{(d-2)\Omega_{d-2}}{16\pi}r_{+}^{d-1}d\frac{1}{l^{2}}+\Phi dQ,
\end{equation}
where $T=\frac{\partial_{r_{+}} f}{4\pi}$ is the temperature, $S=\frac{\Omega_{d-2}}{4}r_{+}^{d-2}$ is the entropy, $\Phi=\sqrt{\frac{d-2}{2(d-3)}}\frac{q}{r_{+}^{d-3}}$ is the electric potential. If we introduce a new ``charge" as in Ref.\cite{Tian:2014goa,Tian2018hlw}
\begin{equation}
  \epsilon=\Omega_{d-2}k^{\frac{d-2}{2}},
\end{equation}
then its conjugate potential is obtained as $\omega=\frac{1}{8\pi}k^{\frac{4-d}{2}}r_{+}^{d-3}$. If we define the black hole volume as $V=\frac{\Omega_{d-2}}{d-1}r_{+}^{d-1}$, then its conjugate pressure is naturally arisen as $P=\frac{(d-1)(d-2)}{16\pi l^{2}}$, and the black hole mass is naturally viewed as enthalpy instead of energy. Finally, the extended first law is obtained as
\begin{equation}\label{firstlaw}
  dM=TdS+\omega d\epsilon+V dP+\Phi dQ.
\end{equation}

\section{the specific heat and phase transition}
\label{sec3}

Hereafter, the investigation will be limited in $d=4$ dimensional space-time and in canonical ensemble with fixed electric charge and topological charge, leaving  other situations for further study. First of all, we would like to analyse the behavior of the specific heat and the related possible phase transition phenomena. The first law can be rewritten in terms of energy $E=M-P V$,
\begin{equation}
  dE=TdS-P dV.
\end{equation}
So the isobaric specific heat can be written as
\begin{eqnarray}
  C_{P,Q,\epsilon}&=&T(\frac{\partial S}{\partial T})_{P,Q,\epsilon}\nonumber\\
  &=&\frac{2\pi r_{+}^{2}(32\pi^{2} P r_{+}^{4}+\epsilon r_{+}^{2}-4\pi Q^{2})}{32\pi^{2} P r_{+}^{4}-\epsilon r_{+}^{2}+12\pi Q^{2}}.
\end{eqnarray}
Since we are in canonical ensemble, $C_{P,Q,\epsilon}$ can be abbreviated as $C_{P}$. From the denominator, we can conclude\\
(1) when $P<\frac{\epsilon^{2}}{1536\pi^{3}Q^{2}}$, $C_{P}$ has two diverge points at
\begin{eqnarray}
  r_{+(1,2)}=\frac{1}{8\pi}\sqrt{\frac{\epsilon \pm \sqrt{\epsilon^{2}-1536(\pi)^{3}Q^{2}P}}{P}},
\end{eqnarray}
which signals a phase transition.\\
(2)when $P=P_{c}=\frac{\epsilon^{2}}{1536\pi^{3}Q^{2}}$,  the two diverge points of $C_{P}$ merge into one at
\begin{eqnarray}
  r_{+}=r_{c}=2\sqrt{\frac{6\pi}{\epsilon}}Q,
\end{eqnarray}
which is the phase transition critical point. The critical temperature $T_{c}=\frac{\epsilon^{3/2}}{24\sqrt{6}(\pi)^{5/2}Q}$.\\
(3)when $P>\frac{\epsilon^{2}}{1536\pi^{3}Q^{2}}$, $C_{P}$ is always larger than zero, so there is no phase transition.

Comparing with the van der Waals equation, the specific volume is defined as\cite{2012JHEP07033K}
\begin{equation}
  v=2r_{+}.
\end{equation}
At the critical point, we obtain an interesting relation
\begin{equation}
  \frac{P_{c}v_{c}}{T_{c}}=\frac{3}{8},
\end{equation}
which is exactly the same as for the van der Waals fluid and RN-AdS black holes. Note that this number which seems to be universal, doesn't depend on the topological charge or electric charge.

All the physical quantities can be rescaled by those at the critical point. Defining
\begin{eqnarray}\label{rsc}
  r_{+}=\tilde{r}r_{c},\,\,\,\,P=\tilde{P}P_{c},
\end{eqnarray}
the isobaric specific heat becomes
\begin{eqnarray}
  C_{P}=\frac{16\pi^{2}Q^{2}}{\epsilon}\frac{\tilde{r}^{2}(3\tilde{P}\tilde{r}^{4}+6\tilde{r}^{2}-1)}{\tilde{P}\tilde{r}^{4}-2\tilde{r}^{2}+1}
  \equiv \frac{16\pi^{2}Q^{2}}{\epsilon} \tilde{C}_{\tilde{P}}
\end{eqnarray}

\begin{figure}
\begin{center}
\includegraphics[scale=0.40]{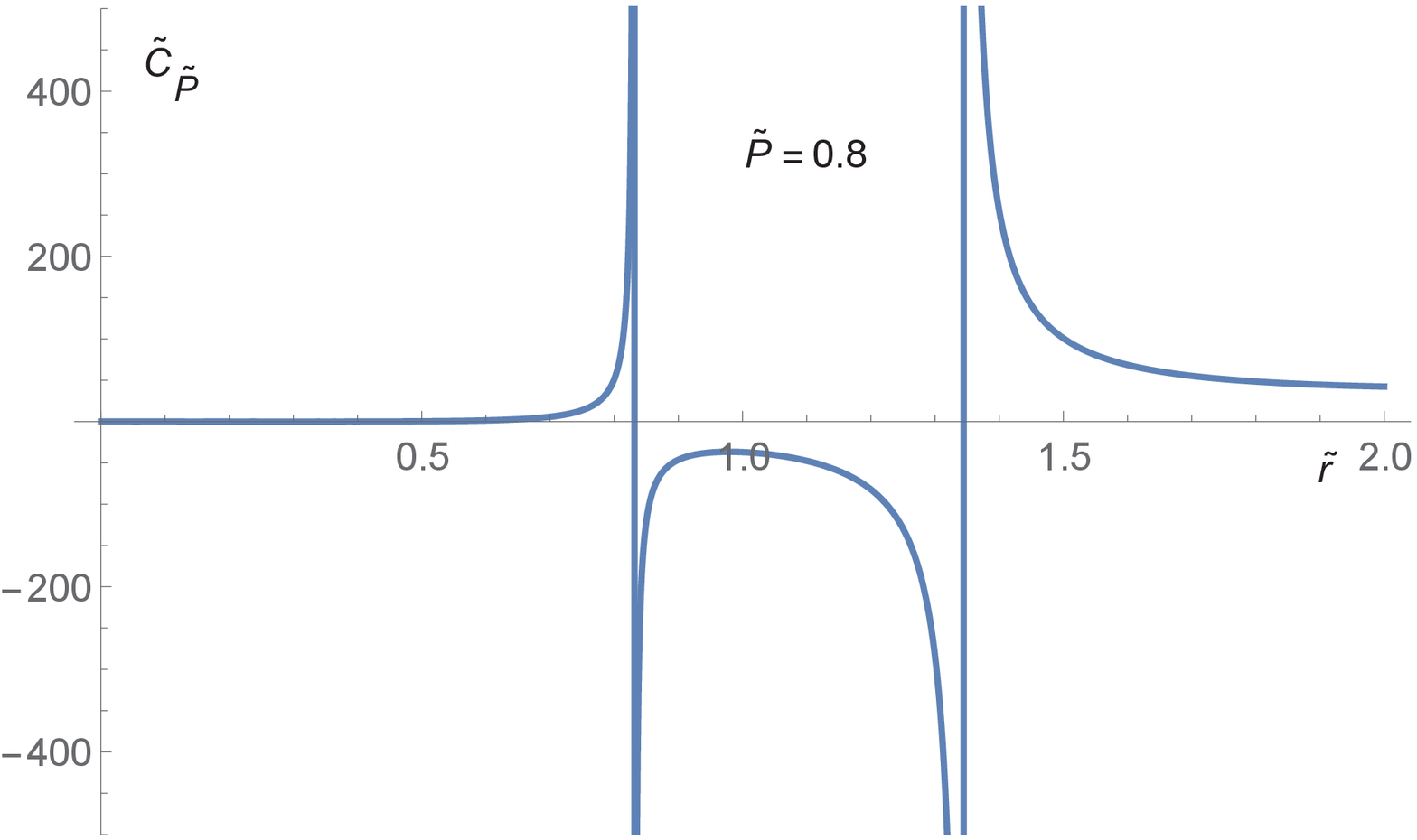}
\includegraphics[scale=0.40]{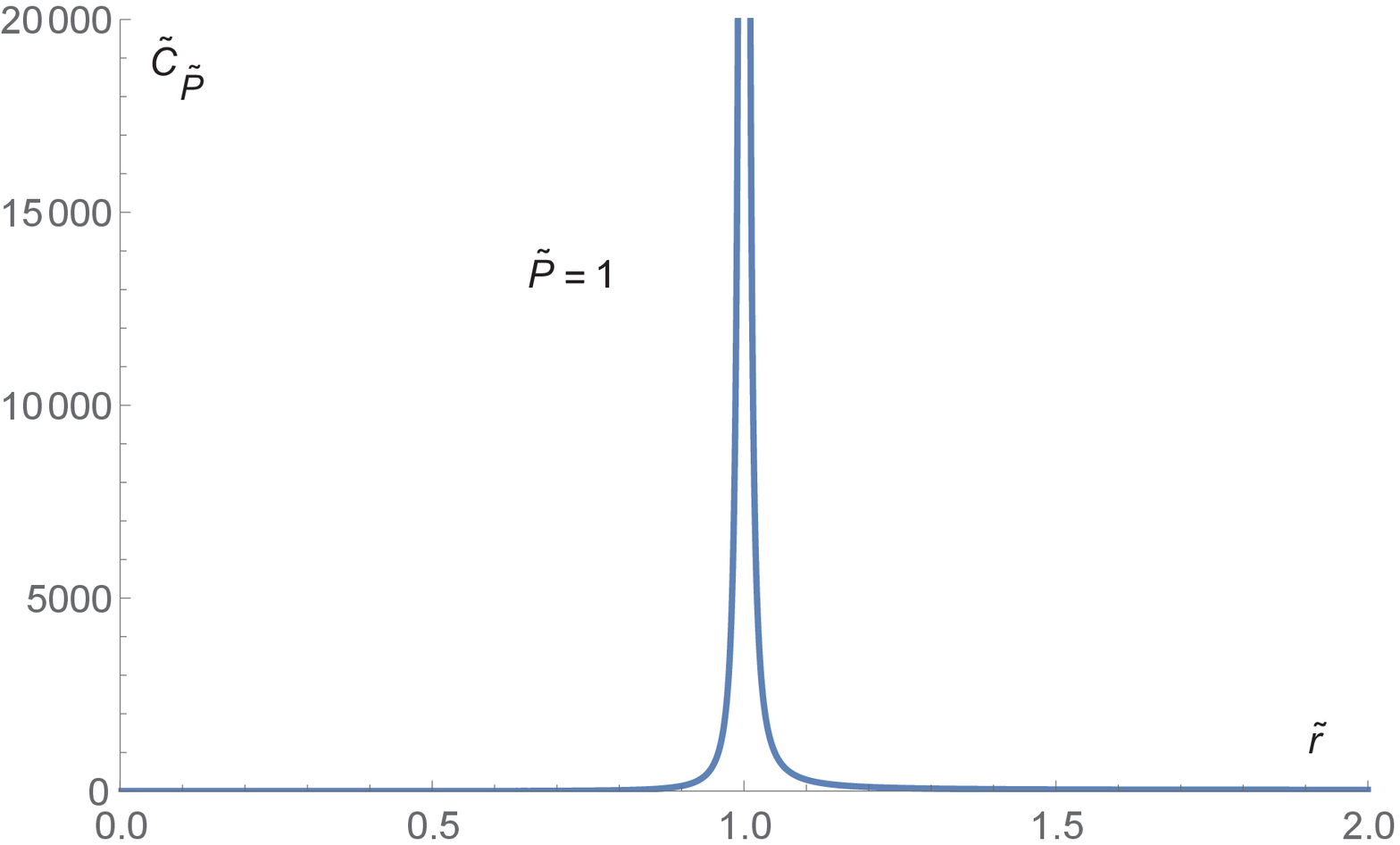}
\includegraphics[scale=0.4]{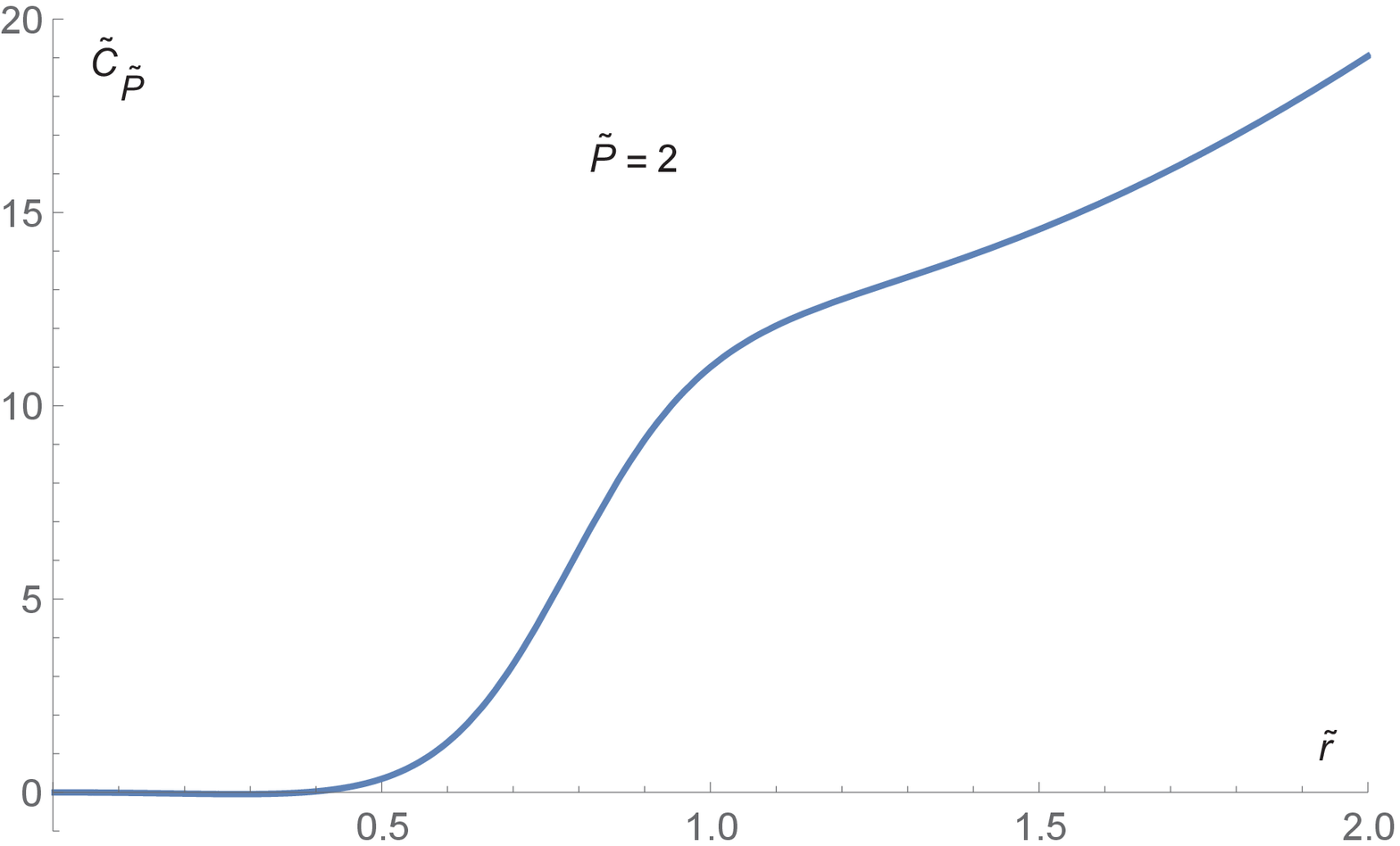}
\end{center}
\caption{The specific heat $\tilde{C}_{\tilde{P}}$ vs. $\tilde{r}$ for $\tilde{P}=0.8<\tilde{P}_{c}$ which has two divergent points,$\tilde{P}=1=\tilde{P}_{c}$ which has only one divergent point and $\tilde{P}=2>\tilde{P}_{c}$ which has no divergent point. }\label{specificheat}
\end{figure}

The behavior of the rescaled specific heat $\tilde{C}_{\tilde{P}}$ for the cases $P<P_{c}$,$P=P_{c}$,$P>P_{c}$ are shown in Fig.\ref{specificheat}. The curve of specific heat for $P<P_{c}$ has two divergent points which divide the region into three parts. Both the large radius region and the small radius region are thermodynamically stable with positive specific heat, while the medium radius region is unstable with negative specific heat. So there is a phase transition which takes place between small black hole and large black hole. The curve of specific heat for $P=P_{c}$ has only one divergent point and always positive denoting that $\epsilon_{c}$ is exactly the critical point. The curve of specific heat for $P>P_{c}$ has no divergent point and always positive, implying the black holes are stable and no phase transition will take place. This behavior of specific heat is very similar to that of the liquid-gas var der Waals system.

\section{oscillatory behavior in $T-S$ and $P-V$ graphs, phase transition coexistence curve }
\label{sec4}

In the last section, we have determined the critical point and found a phase transition when $P\leq P_{c}$. In this section and in the next section, we will derive the analytical phase transition coexistence curve by using different methods.

\subsection{Maxwell equal area law in $T-S$ graph and phase transition coexistence curve }

The temperature and entropy are
\begin{eqnarray}
  T&=&\frac{f'(r_{+})}{4\pi}=\frac{1}{4\pi r_{+}}(k-\frac{q^{2}}{r_{+}^{2}}+\frac{3r_{+}^{2}}{l^{2}})=\frac{\epsilon}{16\pi^{2} r_{+}}-\frac{Q^{2}}{4\pi r_{+}^{3}}+2P r_{+},\nonumber\\
  S&=&\pi r_{+}^{2},
\end{eqnarray}
which can be rescaled by Eq.(\ref{rsc}) to be
\begin{eqnarray}
  T&=&\frac{3\tilde{P}\tilde{r}^{4}+6\tilde{r}^{2}-1}{8\tilde{r}^{3}}\frac{\epsilon^{3/2}}{24\sqrt{6}\pi^{5/2}Q}=\tilde{T}T_{c},\nonumber\\
  S&=&\tilde{r}^{2}\frac{24\pi^{2}Q^{2}}{\epsilon}=\tilde{S}S_{c},
\end{eqnarray}
here $S_{c}\equiv \pi r_{c}^{2}=\frac{24\pi^{2}Q^{2}}{\epsilon}$.
Thus we obtain
\begin{eqnarray}
  \tilde{T}=\frac{3\tilde{P}\tilde{S}^{2}+6\tilde{S}-1}{8\tilde{S}^{3/2}}.
\end{eqnarray}

\begin{figure*}
\includegraphics[scale=0.6]{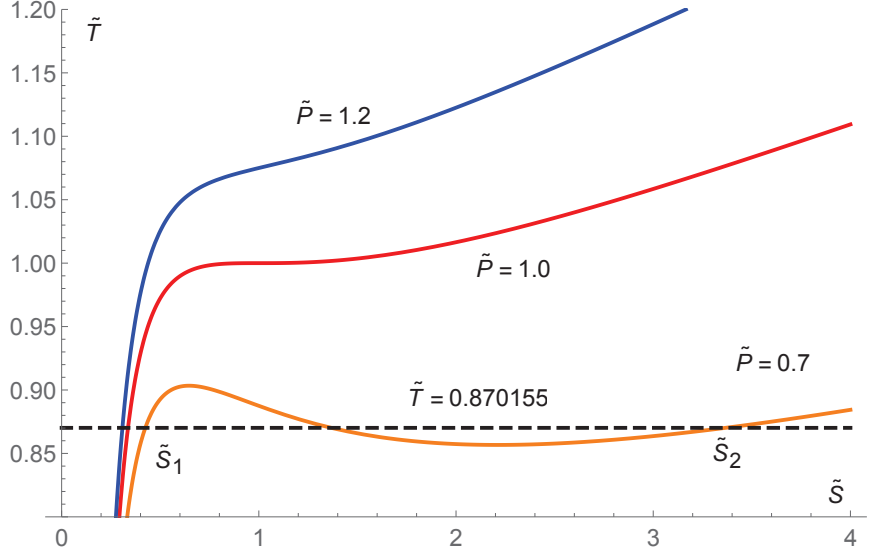}
\caption{$\tilde{T}$ vs. $\tilde{S}$ for $\tilde{P}=0.7,1.0,1.2$. The phase transition take place for $\tilde{P}\leq 1.0$. The dashed black line $\tilde{T}=0.870155$ equally separate the oscillatory part. According to the Maxwell's equal area law, the phase transition point is $(\tilde{T}=0.870155, \tilde{P}=0.7).$ }\label{tsm}
\end{figure*}

From the above equation, we can plot the curve $ \tilde{T}( \tilde{S})$ for different $\tilde{P}$ in Fig.\ref{tsm}. One can see that for pressure $\tilde{P}\leq 1.0$, temperature $ \tilde{T}( \tilde{S})$ curves show  interesting var der Waals system's oscillatory behavior which denote the existence of phase transition. The oscillatory part needs to be replaced by an isobar (denote as $\tilde{T}^{*}$) such that the areas above and below it are equal to each other. This treatment obeys Maxwell's equal area law. In what follows, we will analytically determine this isobar $\tilde{T}^{*}$ for different $\tilde{P}$.

\begin{figure}
\begin{center}
\includegraphics[scale=0.4]{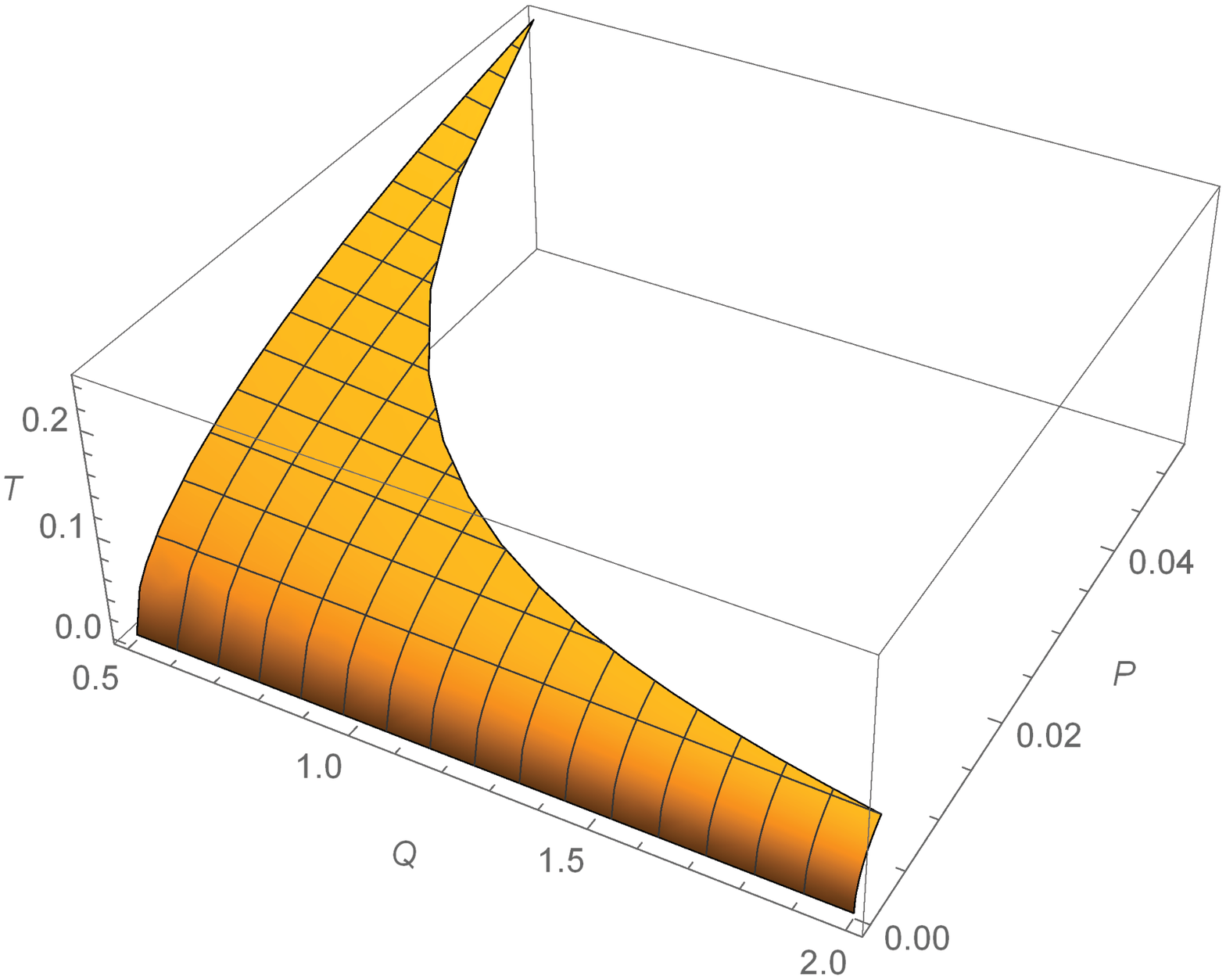}
\includegraphics[scale=0.40]{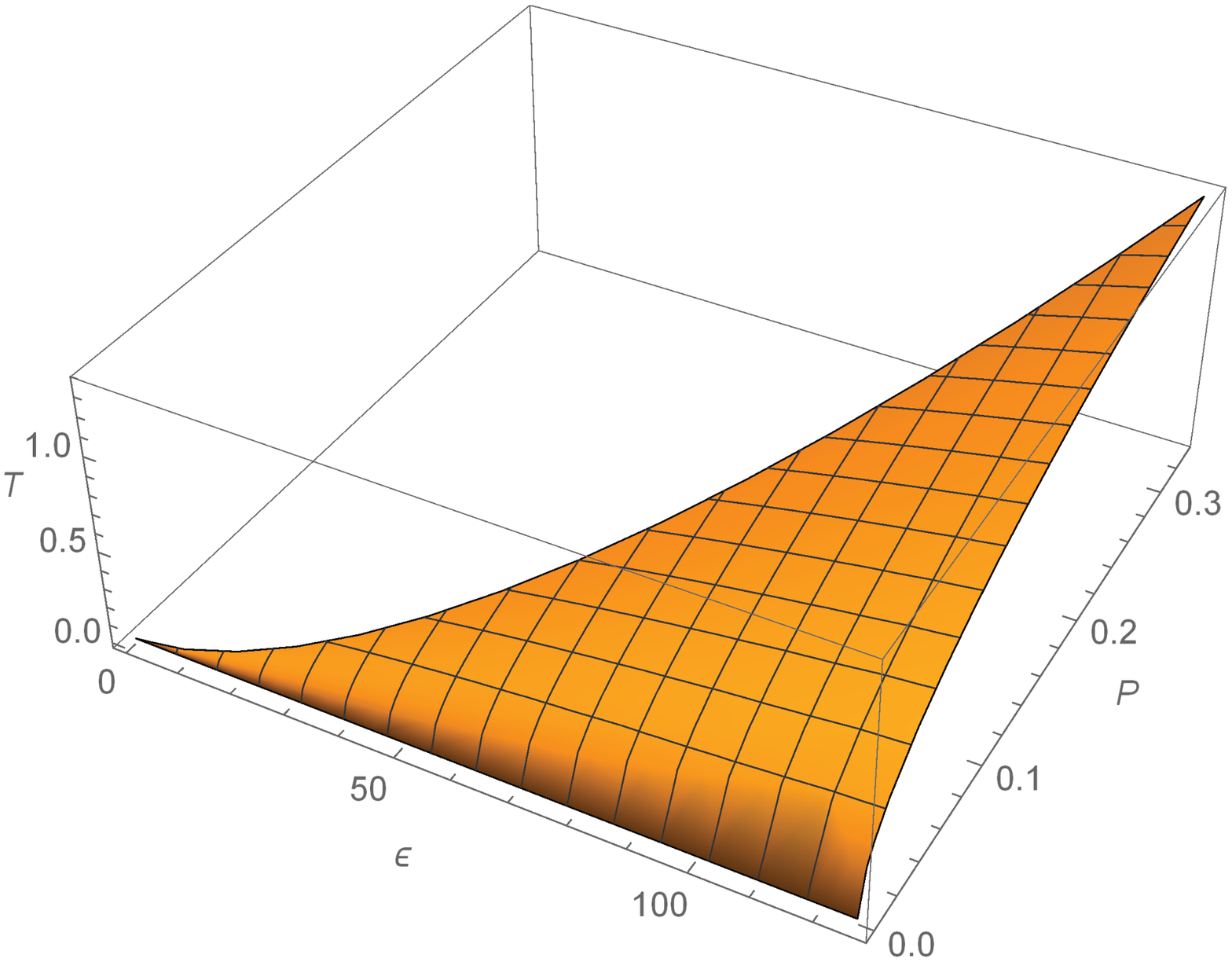}
\end{center}
\caption{The phase transition coexistence curves for $\epsilon=8\pi$ (left) and $Q=1$ (right). The end points are the critical points. }\label{phasetp}
\end{figure}

The Maxwell's equal area law is manifest as
\begin{eqnarray}
  \tilde{T}^{*}(\tilde{S}_{2}-\tilde{S}_{1})&=&\int_{\tilde{S}_{1}}^{\tilde{S}_{2}}\tilde{T}(\tilde{S},\tilde{P})d\tilde{S}\nonumber\\
  &=&\frac{\tilde{P}}{4}(\tilde{S}_{2}^{3/2}-\tilde{S}_{1}^{3/2})+\frac{3}{2}(\tilde{S}_{2}^{1/2}-\tilde{S}_{1}^{1/2})+\frac{1}{4}(\tilde{S}_{2}^{-1/2}-\tilde{S}_{1}^{-1/2}).
\end{eqnarray}
At points $(\tilde{S}_{1},\tilde{T}^{*})$,$(\tilde{S}_{2},\tilde{T}^{*})$, we have two equations
\begin{eqnarray}
  \tilde{T}^{*}&=&\frac{3\tilde{P}\tilde{S}_{1}^{2}+6\tilde{S}_{1}-1}{8\tilde{S}_{1}^{3/2}},\nonumber\\
  \tilde{T}^{*}&=&\frac{3\tilde{P}\tilde{S}_{2}^{2}+6\tilde{S}_{2}-1}{8\tilde{S}_{2}^{3/2}}.
\end{eqnarray}
The above three equations can be solved as
\begin{eqnarray}
  \tilde{S}_{1}&=&\frac{(\sqrt{3-\sqrt{\tilde{P}}}-\sqrt{3-3\sqrt{\tilde{P}}})^{2}}{2\tilde{P}},\nonumber\\
  \tilde{S}_{2}&=&\frac{(\sqrt{3-\sqrt{\tilde{P}}}+\sqrt{3-3\sqrt{\tilde{P}}})^{2}}{2\tilde{P}},\nonumber\\
  \tilde{T}^{*}&=&\sqrt{\tilde{P}(3-\sqrt{\tilde{P}})/2}.\label{tpsr}
\end{eqnarray}
The last equation $\tilde{T}^{*}(\tilde{P})$ is the rescaled phase transition coexistence curve. Then we can make a backward rescale to obtain the phase transition coexistence curve,
\begin{eqnarray}\label{ptcc}
  T=\frac{\sqrt{2 P(3\epsilon-16\sqrt{6}\pi^{3/2}Q\sqrt{P})}}{3\pi}.
\end{eqnarray}
Note that the phase diagram is four dimensional ($T,P,Q,\epsilon$). The condition for the phase transition is that $\epsilon>16\sqrt{6}\pi^{3/2}Q\sqrt{P}/3$. When the topological charge $\epsilon= 0$, there will be no phase transition. While when the electric charge $Q\rightarrow 0$ ($Q>0$), there will be phase transition with the critical temperature $T_{c}\rightarrow \infty$ and pressure $P_{c}\rightarrow \infty$. The detailed dependence of the phase transition on the topological charge can be seen in Eq.(\ref{ptcc}) and in right graph of Fig.\ref{phasetp}.

The phase transition coexistence curves are plotted in Fig.\ref{phasetp} for $\epsilon=8\pi$ (left) and $Q=1$ (right). The end points in the graphs are the critical points. With fixed $\epsilon$ and $Q$ , the phase transition coexistence curve $T(P)$ is reminiscent of the var der Waals system's liquid-gas phase transition coexistence curve.

\subsection{Maxwell equal area law in $P-V$ graph and phase transition coexistence curve}

The pressure and volume are
\begin{eqnarray}
  P&=&\frac{T}{2r_{+}}-\frac{\epsilon}{32\pi^{2}r_{+}^{2}}+\frac{Q^{2}}{8\pi r_{+}^{4}},\nonumber\\
  V&=&\frac{4}{3}\pi r_{+}^{3},
\end{eqnarray}
which can be rescaled to be
\begin{eqnarray}
  P&=&(\frac{8\tilde{T}}{3\tilde{r}}-\frac{2}{\tilde{r}^{2}}+\frac{1}{3\tilde{r}^{4}})\frac{\epsilon^{2}}{1536\pi^{3}Q^{2}}=\tilde{P}P_{c},\nonumber\\
  V&=&\tilde{r}^{3}64\sqrt{6}\pi^{5/2}(\frac{Q^{2}}{\epsilon})^{3/2}=\tilde{V}V_{c}.
\end{eqnarray}
Thus we obtain
\begin{eqnarray}
  \tilde{P}=\frac{8\tilde{T}}{3}\tilde{V}^{-1/3}-2\tilde{V}^{-2/3}+\frac{1}{3}\tilde{V}^{-4/3}.
\end{eqnarray}

\begin{figure}
\begin{center}
\includegraphics[scale=0.6]{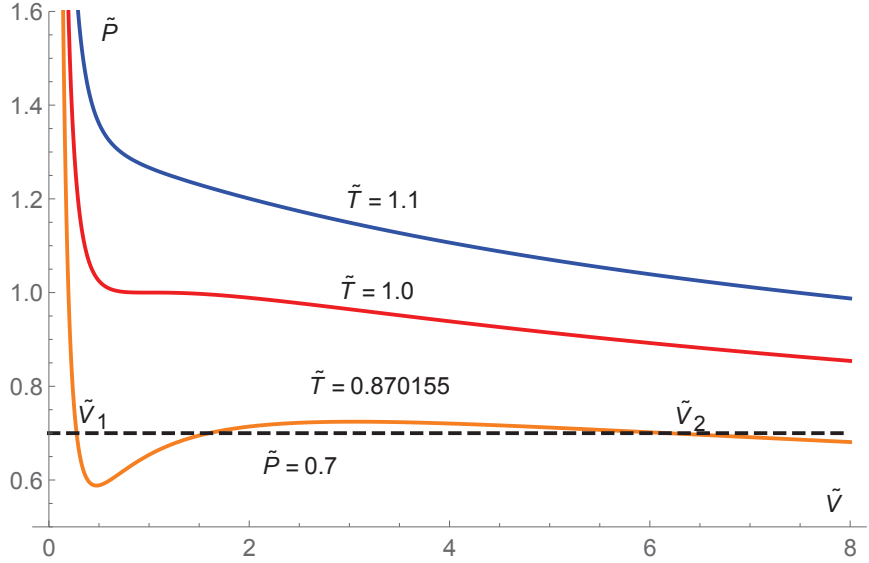}
\end{center}
\caption{$\tilde{P}$ vs. $\tilde{V}$ for $\tilde{T}=0.870155,1.0,1.1$. The phase transition take place for $\tilde{T}\leq 1.0$. The dashed black line $\tilde{P}=0.7$ equally separate the oscillatory part. According to the Maxwell's equal area law, the phase transition point is $(\tilde{T}=0.870155, \tilde{P}=0.7).$ }\label{pvm}
\end{figure}

From the above equation, we can plot the curve $ \tilde{P}( \tilde{V})$ for different $\tilde{T}$ in Fig.\ref{pvm}. One can see that for temperature $\tilde{T}\leq 1.0$, pressure $ \tilde{P}( \tilde{V})$ curves show  interesting var der Waals system's oscillatory behavior which corresponds to the phase transition. Similarly, the oscillatory part needs to be replaced by an isobar (denote as $\tilde{P}^{*}$) such that the areas above and below it are equal to each other. This treatment follows Maxwell's equal area law. The analytical phase transition curve is derived as follows.

The Maxwell's equal area law is manifest as
\begin{eqnarray}
   \tilde{P}^{*}(\tilde{V}_{2}-\tilde{V}_{1})&=&\int_{\tilde{V}_{1}}^{\tilde{V}_{2}}\tilde{P}(\tilde{V},\tilde{T})d\tilde{V}\nonumber\\
  &=&4\tilde{T}(\tilde{V}_{2}^{2/3}-\tilde{V}_{1}^{2/3})-6(\tilde{V}_{2}^{1/3}-\tilde{V}_{1}^{1/3})-\frac{1}{3}(\tilde{V}_{2}^{-1/3}-\tilde{V}_{1}^{-1/3}).
\end{eqnarray}
At points $(\tilde{V}_{1},\tilde{P}^{*})$,$(\tilde{V}_{2},\tilde{P}^{*})$, we have two equations
\begin{eqnarray}
  \tilde{P}^{*}&=&\frac{8\tilde{T}}{3}\tilde{V}_{1}^{-1/3}-2\tilde{V}_{1}^{-2/3}+\frac{1}{3}\tilde{V}_{1}^{-4/3},\nonumber\\
  \tilde{P}^{*}&=&\frac{8\tilde{T}}{3}\tilde{V}_{2}^{-1/3}-2\tilde{V}_{2}^{-2/3}+\frac{1}{3}\tilde{V}_{2}^{-4/3}.
\end{eqnarray}
The above three equations can be solved as
\begin{eqnarray}
  &\,&\tilde{V}_{1}=(\frac{2cos^{2}\varphi}{\tilde{T}}-\sqrt{\frac{4cos^{4}\varphi}{\tilde{T}^{2}}-\frac{\sqrt{2}cos\varphi}{\tilde{T}}})^{3},\nonumber\\
  &\,&\tilde{V}_{2}=(\frac{2cos^{2}\varphi}{\tilde{T}}+\sqrt{\frac{4cos^{4}\varphi}{\tilde{T}^{2}}-\frac{\sqrt{2}cos\varphi}{\tilde{T}}})^{3},\nonumber\\
  &\,&\varphi=\frac{\pi-\theta}{3},\,\,\,\,\,cos\theta=\frac{\sqrt{2}}{2}\tilde{T}\nonumber\\
  &\,&\tilde{P}^{*}=(4cos\frac{\theta}{3}cos\frac{\pi+\theta}{3})^{2}=(1-2cos\frac{arccos(1-\tilde{T}^{2})+\pi}{3})^{2}.
\end{eqnarray}
The last equation $\tilde{P}^{*}(\tilde{T})$ is the rescaled phase transition coexistence curve, and it can be rewritten as
\begin{eqnarray}
  \tilde{T}=\sqrt{\tilde{P}^{*}(3-\sqrt{\tilde{P}^{*}})/2},
\end{eqnarray}
which is exactly same with Eq.(\ref{tpsr}). So the phase transition coexistence curves obtained by applying the Maxwell equal area law in $T-S$ graph and in $P-V$ graph are consistent with each other.

\section{swallow tail behavior in $G-T$ and $G-P$ graphs, phase transition coexistence curve}
\label{sec5}

In Sec.\ref{sec2}, we see that  the black hole mass can be interpreted as enthalpy. Thus the Gibbs free energy is
\begin{eqnarray}\label{gibbs}
  G=M-TS,
\end{eqnarray}
and its differential form in canonical ensemble can be obtained from Eq.(\ref{firstlaw})
\begin{eqnarray}
  dG=-SdT+\omega d\epsilon+V dP+\Phi dQ=-SdT+V dP,
\end{eqnarray}
which denotes the Gibbs free energy is a function of temperature and pressure.

\begin{figure}
\begin{center}
\includegraphics[scale=0.4]{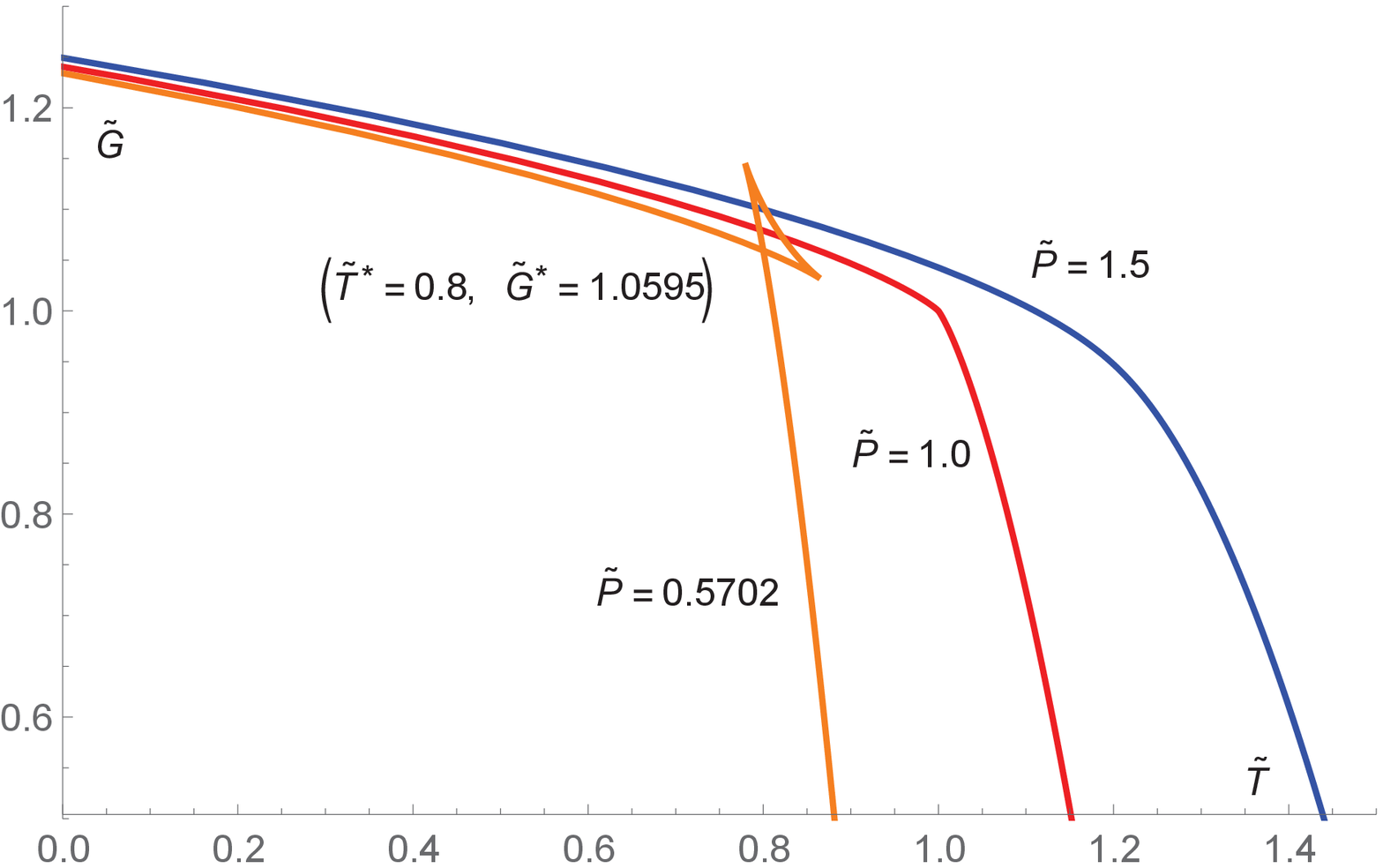}
\includegraphics[scale=0.4]{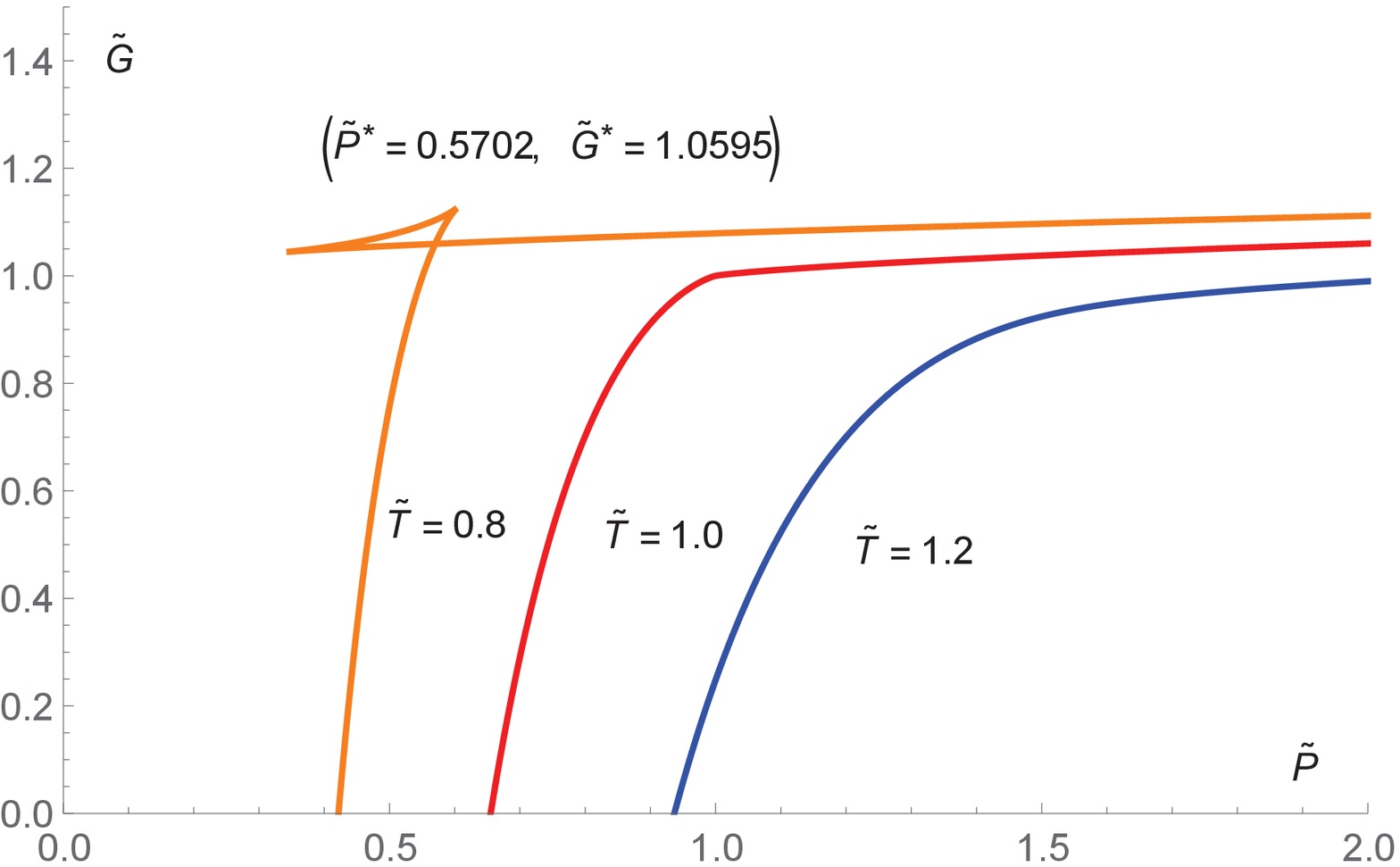}
\end{center}
\caption{Left panel shows $\tilde{G}$ vs. $\tilde{T}$ for different $\tilde{P}=0.5702,1.0,1.5$. When $\tilde{P}\leq 1.0$, there is a var der Waals system's swallow tail behavior, and the cross point is the  phase transition point.  Right panel shows $\tilde{G}$ vs. $\tilde{P}$ for different $\tilde{T}=0.8,1.0,1.2$. When $\tilde{T}\leq 1.0$, there is also a var der Waals system's swallow tail behavior, and the cross point is the  phase transition point. }\label{gtp}
\end{figure}

Substituting black hole mass, temperature and entropy into Eq.(\ref{gibbs}), then making a rescaling by the quantities at the critical point, we will obtain
\begin{eqnarray}\label{gten}
  &\,&G=\frac{\epsilon}{16\pi}r_{+}-\frac{2\pi}{3}P r_{+}^{3}+\frac{3Q^{2}}{4r_{+}}=\frac{3+6\tilde{r}^{2}-\tilde{P}\tilde{r}^{4}}{8\tilde{r}}\sqrt{\frac{\epsilon Q^{2}}{6\pi}}=\tilde{G}G_{c},\nonumber\\
  &\,&T=\frac{\epsilon}{16\pi^{2} r_{+}}-\frac{Q^{2}}{4\pi r_{+}^{3}}+2P r_{+}=\frac{3\tilde{P}\tilde{r}^{4}+6\tilde{r}^{2}-1}{8\tilde{r}^{3}}\frac{\epsilon^{3/2}}{24\sqrt{6}\pi^{5/2}Q}=\tilde{T}T_{c}
\end{eqnarray}
From the above equation, we can plot $\tilde{G}-\tilde{T}$ curves for different $\tilde{P}$, and plot $\tilde{G}-\tilde{P}$ curves for different $\tilde{T}$ in Fig.\ref{gtp}.
One can see that both panels display a var der Waals system's swallow tail behavior when $\tilde{P}\leq 1.0$ and $\tilde{T}\leq 1.0$. Note that $\tilde{G}(\tilde{T})$ doesn't depends on the topological charge or the electric charge. But by applying a backward rescale, one can find that the swallow tail curves of $G(T)$ in Eq.(\ref{gten}) for different $\epsilon$ and $Q$ have conformal symmetry.  Since phase transition take place where the system's two phases have equal Gibbs free energy, temperature and pressure, the swallow tail's intersection point is exactly the phase transition point. As a result, left panel and right panel have an duality relation and they are equal to each other.  Then we will only analyse $\tilde{G}-\tilde{T}$ graph to derive the phase transition coexistence curve as follows.

In $\tilde{G}-\tilde{T}$ graph, at the phase transition point($\tilde{T}^{*},\tilde{G}^{*}$) for fixed $\tilde{P}$, we assume that the black hole radius is $\tilde{r}_{1}$ for one phase and $\tilde{r}_{2}>\tilde{r}_{1}$ for the other phase. Thus we will have the following equations from Eq.(\ref{gten}),
\begin{eqnarray}
  &\,&\tilde{G}^{*}=\frac{3+6\tilde{r}_{1}^{2}-\tilde{P}\tilde{r}_{1}^{4}}{8\tilde{r}_{1}}=\frac{3+6\tilde{r}_{2}^{2}-\tilde{P}\tilde{r}_{2}^{4}}{8\tilde{r}_{2}},\nonumber\\
&\,&\tilde{T}^{*}=\frac{3\tilde{P}\tilde{r}_{1}^{4}+6\tilde{r}_{1}^{2}-1}{8\tilde{r}_{1}^{3}}=\frac{3\tilde{P}\tilde{r}_{2}^{4}+6\tilde{r}_{2}^{2}-1}{8\tilde{r}_{2}^{3}}.
\end{eqnarray}
The above equations can be solved as
\begin{eqnarray}
  \tilde{r}_{1}&=&\frac{\sqrt{3-\sqrt{\tilde{P}}}-\sqrt{3-3\sqrt{\tilde{P}}}}{\sqrt{2\tilde{P}}},\nonumber\\
  \tilde{r}_{2}&=&\frac{\sqrt{3-\sqrt{\tilde{P}}}+\sqrt{3-3\sqrt{\tilde{P}}}}{\sqrt{2\tilde{P}}},\nonumber\\
  \tilde{G}^{*}&=&\frac{\sqrt{6-2\sqrt{\tilde{P}}}}{2},\nonumber\\
  \tilde{T}^{*}&=&\sqrt{\tilde{P}(3-\sqrt{\tilde{P}})/2}.
\end{eqnarray}
The last equation is exactly same with Eq.(\ref{tpsr}). So the phase transition coexistence curves obtained by analysing the Gibbs free energy in $G-T$ graph and $G-P$ graph , or by applying the Maxwell equal area law in $T-S$ graph and $P-V$ graph are consistent with each other.

\section{Conclusion and discussion}
\label{sec6}

Treating the cosmological constant as a variable\cite{Kastor:2009wy,2012JHEP07033K} and the spatial curvature as topological charge\cite{Tian:2014goa,Tian2018hlw}, thermodynamics of electrically charged Reissner-Nordstr$\ddot{o}$m AdS black holes are investigated. Firstly, by variation of the equipotential equation on horizon, the extended thermodynamic first law is obtained. From the extended first law, a conjugate potential correspondent to the topological charge is arisen. Meanwhile, if the black hole volume is defined as $V=\frac{\Omega_{d-2}}{d-1}r_{+}^{d-1}$, then its conjugate pressure is naturally assigned as the cosmological constant and the black hole mass as enthalpy.

Secondly, in four dimensional space-time and canonical ensemble with fixed electric charge and topological charge, the isobaric specific heat $C_{P}$ is calculated and the corresponding divergent solutions are derived. The two solutions merge into one at the critical point with $P_{c}=\epsilon^{2}/(1536\pi^{3}Q^{2}), r_{c}=2\sqrt{6\pi/\epsilon}Q$. When $P<P_{c}$, the curve of specific heat has two divergent points and is divided into three regions. The specific heat is positive for both the large radius region and the small radius region which are thermodynamically stable, while it is negative for the medium radius region which is unstable. When $P>P_{c}$, the specific heat is always positive implying the black holes are stable and no phase transition will take place.

Thirdly, rescaling the quantities by those at the critical point, the behavior of temperature in $\tilde{T}-\tilde{S}$ graph and the behavior of pressure in $\tilde{P}-\tilde{V}$ graph are studied. They exhibit the interesting van de Waals gas-liquid system's behavior. When $\tilde{P}>1, \tilde{T}>1$, the curves vary monotonically and no phase transition will take place. When $\tilde{P}<1, \tilde{T}<1$, the curves display an oscillatory behavior which signals phase transition. The oscillatory part is replaced by an isobar according to the Maxwell's equal area law and the analytical phase transition coexistence curves (rescaled) are obtained which are consistent with each other. Then by making a backward rescale, the explicit phase transition coexistence curve is derive in Eq.(\ref{ptcc}) and the phase diagrams are shown in Fig.\ref{phasetp}.

Fourthly, van der Waals system's swallow tail behavior is observed in the $\tilde{G}-\tilde{T}$ graph and $\tilde{G}-\tilde{P}$ graph when $\tilde{P}<1, \tilde{T}<1$. The swallow tail's intersection point is the phase transition point. By analytically solving the constraint equations, the rescaled phase transition coexistence curve is obtained which is consistent with those derived in $\tilde{T}-\tilde{S}$ graph and $\tilde{P}-\tilde{V}$ graph.

From the above detailed study in canonical ensemble, the analogy of RN-AdS black hole as van der Waals system have been examined when the spatial curvature is treated as topological charge and the cosmological constant is treated as pressure. Both the systems share the same oscillatory behavior and swallow tail behavior. Comparing with the case when the spatial curvature is fixed \cite{2012JHEP07033K,2013arXiv13053379S}, our phase transition diagram in Fig.\ref{phasetp} is four dimensional ($T,P,Q,\epsilon$), which is more rich with an extra parameter - the topological charge.

A further investigation in grand canonical ensemble is outside the scope of this paper, but it is surely a very interesting direction for future research. The influence of this topological charge on black hole thermodynamics in other gravity theories ( such as the Lovelock, Gauss-Bonnet theory, f(R) theory ) and different dimensional space-time also deserves to be disclosed in the future research. Another interesting future research line is the comparison of the influence between the electric charge and topological charge.

In the end, we would like to point out that the black hole thermodynamics discussed in this paper is based on the first law derived from the equipotential surface $f(r)=c$ with $c=0$. A more conventional way to compute thermodynamics quantities is the Euclidean formalism, where the free energy is computed firstly, then the other quantities follow from it. The difference between these two formalisms remains unknown which deserve to be investigated in future.

\begin{acknowledgments}
This research is supported by Department of Education of Guangdong Province, China (Grant Nos.2017KQNCX124).
\end{acknowledgments}


\end{document}